# Motor Rotation Speed Estimation based on Magnetic Inductive Sensing

## Rahul Hoskeri and Hua Huang

Department of Computer Science and Engineering, Univeristy of California, Merced
CA, USA

## ABSTRACT

Rotation speed is a key metric for many applications, such as calibrating electric motors in a factory, monitoring a car's engine health, detecting faults in electrical appliances, and more. However, existing measurement techniques have several drawbacks, including the need for line-of-sight visibility, intrusive machine modification, fixed sensor deployment, or short sensing ranges. In this paper, we introduce MagTach, a handheld hardware-software system for rotation speed measurement that leverages electromagnetic field sensing to provide highly accurate and convenient readings. MagTach estimates a device's rotation speed by passively sensing its naturally emitted magnetic field, which demonstrates strong occlusion-penetrating capabilities due to its high skin depth. To overcome the challenge of rapid magnetic signal strength degradation over distance, we custom-designed magnetic inductive sensors with superior sensitivity to time-varying magnetic fields. We developed a spatial-filtering-based algorithm that significantly enhances the signal-to-noise ratio by combining measurements from multiple sensors. We developed the novel Pyramid Power Spectrum Parsing Network (PPSP Net), which detects motor magnetic signals under severe noise, and a fuzzy-logic-based algorithm that delivers highly accurate rotation speed estimations despite uncertain signal detection results. Through extensive evaluations, MagTach demonstrated state-of-the-art performance, achieving mean estimation errors of 0.21%, 0.34%, and 0.54% at distances of 50cm, 70cm, and 100cm, respectively. It remained robust across varying device orientations, operated reliably at speeds between 650 and 9000 RPM, and generalized well to previously unseen devices. MagTach also maintained high monitoring accuracy despite the presence of occluding objects, such as a 15cm-thick wall and 0.8mm-thick stainless steel sheets, all while featuring a compact handheld form factor with low power consumption of 760mW.

## 1 INTRODUCTION

Electric machines, i.e., devices that convert electric power into rotational movements, are ubiquitous in our daily lives and play crucial roles in various industries, including energy, aviation, automotive, and home appliances. In automotive vehicles, the rotation speed of the engine or motor is a crucial operating parameter that is continuously monitored in real-time. In many precision medical devices, such as insulin pumps, precise monitoring of the pump is crucial to ensuring the correct dosage of medicine is delivered.

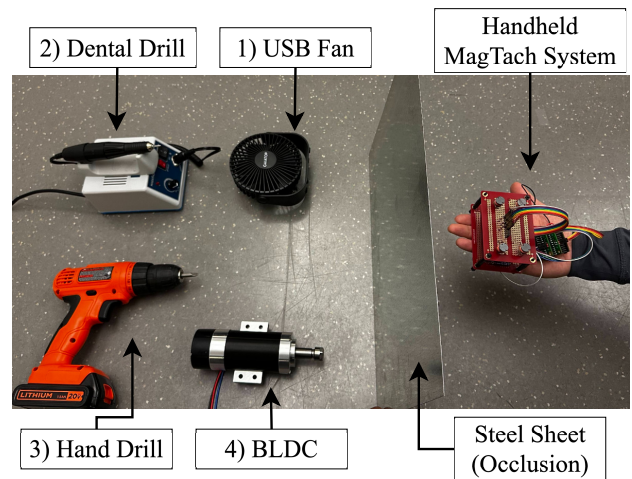

**Figure 1: MagTach Setup**

Various sensing technologies have been employed to determine rotational speeds. Laser tachometers are low-cost ( ≈ \$20), offer high monitoring accuracy (error ≤ 0.4%), and have a reasonable sensing range (≈ 20cm). As a result, they have achieved widespread adoption [52]. However, they require attaching reflective tapes to the rotating blades of devices and maintaining Line-of-Sight (LoS) visibility, which can be impractical for many applications. Furthermore, while laser tachometers are designed as portable devices, it is challenging for users to aim at the extremely small label on a rotating target, especially when the object is vibrating or the measurement distance is long. Additionally, the accuracy of laser tachometers degrades significantly when used in handheld mode [56]. RF-based systems can achieve non-intrusive sensing of rotating devices without any modification [19, 13, 14]. However, due to the shallow skin depth of the selected radio signals, they often have limited capabilities for non-LoS sensing. Additionally, these RF-based systems require fixed setups, partly because of their sensitivity to Doppler effects caused by hand instability, making them unsuitable for hand-held usage [26]. To eliminate the line-of-sight visibility requirements, there exist magnetic sensing based tachometers, such as the Hall-effect tachometers [42, 18, 34, 1, 49, 2, 43]. However, due to the low-frequency magnetic fields' severe path loss, these magnetic sensors can operate only at very close distances ($\leq 5cm$) or require intrusive installation [19], making them usable only for specialized devices, such as vehicles and industrial instruments. A rotation speed sensing solution that offers a reasonable sensing range, is non-intrusive and hand-held, and does not require line-of-sight (LoS) visibility still remains a missing gap in literature.

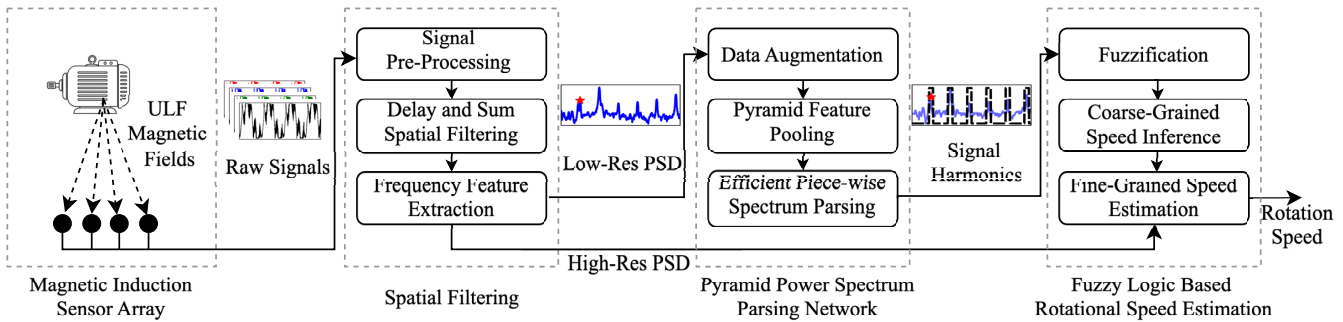

**Figure 2: MagTach System Overview**

We introduce MagTach, illustrated in Figure 1, a hand-held rotation speed monitoring system that measures rotation speeds by detecting a device's emitted magnetic field. The low-frequency motor magnetic field demonstrates exceptional capabilities for sensing through obstacles, such as the 0.8mm-thick steel sheet illustrated in Figure 1, due to its high skin depth. Additionally, the system is passive, enabling a compact form factor and low energy consumption. Through a combination of specially designed hardware and algorithms, MagTach achieves over 10x greater sensing range compared to existing magnetic-sensing-based solutions [42, 18, 34, 1, 49, 2, 43] while maintaining accuracy during hand-held use.

The motor's magnetic signal experiences severe path loss, with its power inversely proportional to the fourth power (or higher) of the distance ($p \propto 1/d^4$) [40]. In contrast, background noise remains nearly constant in the environment, resulting in a diminishing signal-to-noise ratio and posing a fundamental challenge in monitoring the rotation speed. The key observation that enables MagTach is that the motor's emitted magnetic signal exhibits not only a fundamental frequency that matches the rotation speed, but also significant harmonic frequencies that are integer multiples of the fundamental frequency. Different from the existing magnetic tachometers that focus on detecting the fundamental frequency from the signal, we instead seek to sense, detect, and interpret the signal harmonics across a wide band of the signal spectrum, and combine the detection results to estimate the motor rotation speeds. Our contributions are summarized as follows.

- MagTach is the first electromagnetic-based hardware software sensing system capable of measuring the rotation speed of a device from a distance of up to 1 meter. MagTach's use of magnetic inductive sensing provides unique advantages over other technologies. It can estimate rotation speeds through occlusions and features a lightweight, handheld form factor, enabling versatile deployment configurations.

- To capture the magnetic signals over longer distances, we custom-design magnetic inductive sensors that are highly sensitive to time-varying motor magnetic fields. Additionally, we employ a spatial filtering technique, which utilizes a sensor array and a delay-and-sum algorithm, to further enhance signal quality.

- To detect motor signals in the presence of significant noise, we design the Pyramid Power Spectrum Parsing Network (PPSP Net), which leverages the signal's global context to identify signal harmonics. To address uncertainties caused by background noise and interference, we further develop a fuzzy logic-based algorithm to infer rotation speeds based on uncertain harmonic detection.

- We implemented a prototype system and conducted extensive evaluations. MagTach achieved state-of-the-art performance with mean estimation errors of 0.21%, 0.34%, and 0.54% at distances of 50cm, 70cm, and 100cm, respectively, while remaining robust across various operating conditions, including different rotation speeds, device orientations, occluding obstacles, and device types. Our implementation has a compact form-factor, is energy-efficient, and supports handheld operation.

## 2 CHALLENGES AND SYSTEM OVERVIEW

The principle of MagTach is illustrated in Figure 2, where we use magnetic inductive sensors to measure the magnetic field emissions of the motor.

**Rapid Path Loss of the Magnetic Field.** The low frequency ($\leq 3,000Hz$ in our current implementation) motor

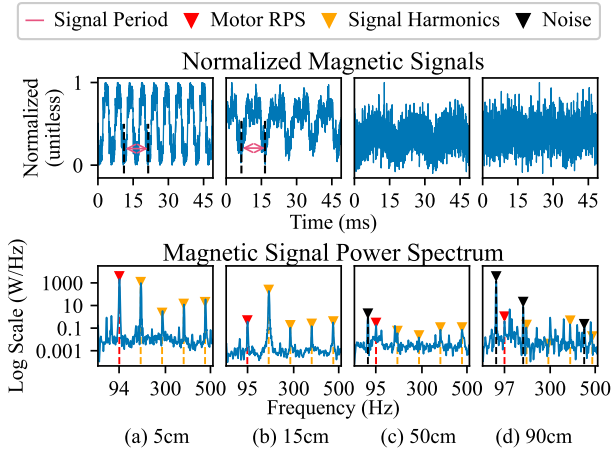

**Figure 3: BLDC Magnetic Signals at Different Distances**

magnetic field power rolls off inverse fourth $(1/d^4)$ or higher over distance [40], which is significantly faster than the radio signals, such as the WiFi [13], millimeter-wave [14], and OAM wave [19] that have signal power rolls off as the inverse square of the distance $(1/d^2)$. As illustrated in the second row of Figure 3, we can see that the power spectral density at the rotation frequency (94 Hz) drops from above 1000W/Hz at 5cm (a) to around 1W/Hz and below 0.1W/Hz at 15cm (b) and 50cm (c), respectively, representing over 1,000 fold reduction in signal strength.

To capture the motor magnetic field at a reasonable distance, we propose to use a magnetic inductive sensor array, as illustrated on the left side of Figure 2. As to be discussed soon in Section 3, this design has an improved sensitivity in detecting the motor signal's harmonic components, and leverages the spatial filtering technique to further enhance the signal amplitude.

**Signal Detection under Significant Noises.** Despite the over four orders of magnitude drop in the signal strengths, the noise levels remain approximately unchanged, resulting in a low signal-to-noise ratio at a longer distance, which poses a significant challenge for accurate motor magnetic signal detection. For example, in second row of Figure 3 (c), we can see that the motor signal at 97 Hz is around 100 times weaker than the background noise at 60 Hz. Similarly, in Figure 3 (d), the noise signal is significantly stronger than the motor signal.

When the signal quality is good, we can use classical digital signal processing techniques, such as the auto-correlation and short-time Fourier transform algorithms, for detecting the motor's rotation speeds. However, when the signal-to-noise ratio is low, these signal processing algorithms' performance deteriorates rapidly. To overcome the signal detection challenge, we design a data-driven signal detection algorithm that exploits a unique feature of the motor magnetic signals, that significant harmonic components exist in the signals. As illustrated in the middle of Figure 2, we develop the Power Spectrum Parsing Network (PPSP), whose goal is to detect the magnetic signal harmonics across a wide spectrum. PPSP leverages the global signal context, the signal harmonics, to assist in the signal harmonic detection.

**Rotation Speed Estimation Under Uncertain Signal Detection.** The environmental noises and interferences can still cause detection errors for the PPSP Net, including both false positive and false negative detections, especially when the motor is at a long distance. Our key insight is that the detection errors caused by noises often occur randomly across the frequency spectrum, which are different from the periodic appearance of the signal harmonics.

To overcome the signal detection uncertainties caused by noises, we design a fuzzy-logic-based algorithm (on the right of Figure 2) that firstly aggregates detection results from multiple signal harmonics to roughly estimate the motor speeds, and then uses a high-resolution PSD to obtain a fine-grained rotation speed estimation. Our experiments showed the fuzzy-logic-based algorithm maintains highly accurate detection even when the noises and interferences overwhelm the motor magnetic signals.

## 3 MAGNETIC SIGNAL ACQUISITION AND PROCESSING

### 3.1 The Magnetic Inductive Sensor Design

**The Motor Magnetic Field Sensing Model.** In the previous section, we intuitively observed that the motor magnetic field contains significant harmonic components in the frequency domain. In this subsection, we aim construct a mathematical model for this phenomenon. Let $B(t)$ denote the motor magnetic field at the location of the inductive sensor, which has an axis of $\vec{m}$. Then the magnetic field components along the sensor's direction, denoted by $B_m(t)$ can be calculated in Equation 1.

We then perform the Fourier decomposition on the motor magnetic field $B_m(t)$ in the second line of Equation 1. In this equation, we use $P$ to denote the signal period, and thus $1/P$ for the fundamental frequency, and $k/P, k = 2, ..., N$ for harmonic frequencies. $D_k, k = 2, ..., N$ represents the amplitudes of the signal harmonic components. As discussed in our preliminary observations and reported in literature [8], the values of $D_k, k = 2, ..., N$ are non-zero, and are sometimes comparable to or even greater than $D_1$, which is the motor magnetic fields fundamental frequency component.

$$
\begin{aligned}
B_m(t) &= B(t) \cdot \vec{m} \\
&= B_m(0) + \sum_{k=1}^{N} D_k \cos(\tfrac{2\pi k}{P} t - \phi_k)
\end{aligned}
\tag{1}
$$

Harmonics have been reported in prior rotation speed estimation systems [19, 14]. We have identified two potential reasons for the motors to generate magnetic signal harmonics in the magnetic sensing system. a) Abrupt changes in current. In brushed motors, the rotors periodic rotation generates a reversal of the currents in the motor, which will generate harmonics in the magnetic fields. b) There are multiple poles in the motor. Most motors are consist of multiple ($N_p > 1$) rotating poles. Each time a motor pole rotates to the position close to the sensor, a pulse in the sensor measurement will be generated, which corresponds to a significant harmonic component at the frequency of $N_p/P$.

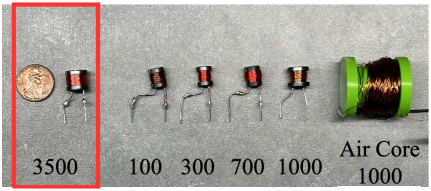

**Figure 4: Inductive Sensing Coils**

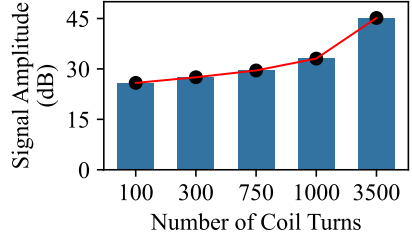

**Figure 5: Impacts of Coil Turns**

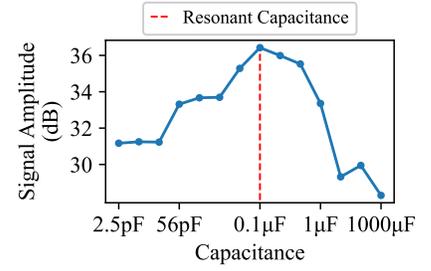

**Figure 6: Impacts of Capacitance**

**Sensor Designs.** To capture the signals' strong harmonic components, we propose to use the magnetic periodic inductive sensors, instead of the popular Hall effect sensors, due to its ability to achieve much higher sensitivities in measuring the magnetic signals. Besides, as reported in [11], the inductive sensors have the potential to capture low-frequency magnetic signals through obstacles, including the wall.

According to the Faraday's law of induction, a time-varying motor magnetic field $B_m(t)$ will induce a voltage $s(t)$ at an inductive sensor, as described in Equation 2. In this equation, we use $\mu_0$, $\mu_r$, $n$, and $A$ to denote the magnetic permeability of empty space, the relative magnetic permeability of the sensor's core, the number of rounds, and the area of the sensor, respectively. From Equation 2, we can see that the amplitude of the $k$th signal harmonic, $D_k \cdot k \cdot \frac{2\pi}{P}$, is linearly proportional to the harmonic multiple $k$. Intuitively, the higher a signal harmonics' component is, the greater multiple $k$ will amplify the signal component.

$$
\begin{aligned}
s(t) &= -\mu_0 \cdot \mu_r \cdot n \cdot A \cdot \frac{dB_m(t)}{dt} \\
&= \mu_0 \cdot \mu_r \cdot n \cdot A \cdot \sum_{k=1}^{N} D_k \cdot k \cdot \frac{2\pi}{P} \sin(\frac{2\pi k}{P}t - \phi_k)
\end{aligned} \quad (2)
$$

We performed experiments to understand how the key design parameters, the number of turns, the coil cross-section area, and the material of the cores, influence the measurements. We selected ferrite cores over air cores, which was adopted in [11], because ferrite cores boost the value of the relative magnetic permeability $\mu_r$ to hundreds from 1, and thus allow for smaller sensor size. As illustrated in Figure 4, we tested induction coils with different turns. As depicted in Figure 5, increasing the number of coil turns enhances the signal amplitude. Notably, a coil with 3500 turns and a sensor area of 314.15, mm$^2$ (highlighted by a red outline in Figure 4) achieved an optimal balance between size and signal quality.

**Impedance Matching.** To further enhance the signal, we adopt the sensing coil resonance enhancement technique described in [23]. A capacitor, $C_q$, is connected in parallel with the inductive sensor. If the signal frequency is $f_m$ and the sensor coil inductance is $L$, the optimal value for $C_q$ can be calculated as $C_q = \frac{1}{4\pi^2 f_m^2 L}$. We tested different capacitors, and the resulting signal amplitudes are shown in Figure 6. As observed, the greatest gain in signal amplitude is achieved when the capacitor is $0.1\mu F$, leading to an impedance match.

## 3.2 Spatial Filtering

To further improve the signal-to-noise ratio, we next explore combining the measurements from multiple sensors. Similar techniques, including the MUSIC (<u>MU</u>ltiple <u>SI</u>gnal <u>C</u>lassification) algorithm, have been used for frequency estimation [10] and radio angle of arrival (AoA) finding [41, 50, 36]. In this project, we performed experiments to verify the assumption that the motor magnetic field measurements from different sensors are correlated with each other by time shifts. We observed that the motor magnetic field has a different propagation pattern from the radio frequency signals studied in the aforementioned work. This observation motivates us to leverage the Spatial Filtering algorithm to significantly enhance the signal quality.

**Preliminary Experiment.** We performed experiments to understand how the motor magnetic signals collected at different locations are related to each other. On a 2D plane, we place the electric motor (a USB fan) at location $(x, y)$. We then use $s_{x,y}^1(t)$ and $s_{x,y}^2(t)$ to denote the sensor measurements collected at locations $(-8, 0)$ and $(8, 0)$ respectively. Based on the sensor measurements, we define the Signal Enhancement Ratio (SER), which quantifies if $s_{x,y}^1(t)$ and $s_{x,y}^2(t)$ are time-shifted versions of each other, as follows.

$$
\text{SER}_{x,y}(\Delta T) = \frac{RMS[s_{x,y}^1(t) + s_{x,y}^2(t - \Delta T)]}{RMS[s_{x,y}^1(t)] + RMS[s_{x,y}^2(t)]}. \quad (3)
$$

$\text{SER}_{x,y}(\Delta T)$ has a large value of close to 1 when $s_{x,y}^1(t)$ equals $s_{x,y}^2(t - \Delta T)$, and $\text{SER}_{x,y}(\Delta T)$ will be close to zero if the two signals are uncorrelated. We next move the electric motor on the 2D plane within the area of $[-8cm, 8cm] \times [0cm, 15cm]$ with 1cm increments. We calculate the values of $SER_{x,y}(\Delta T)$ with different time delays $\Delta T$ in Figure 7.

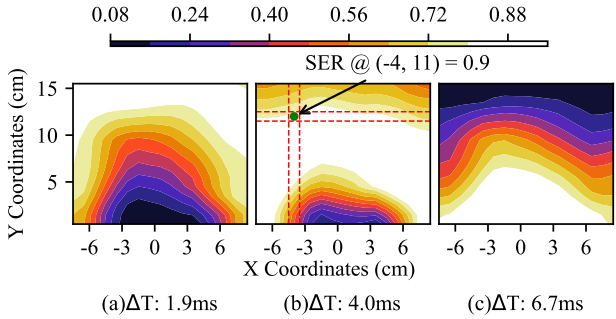

**Figure 7: Signal Enhancement Ratio**

From these results, we can see that the hypothesis is validated. For example, as highlighted by red dotted lines in Figure 7 (b), we can see that the SER value has a value of 0.9 at the location $(-4, 11)$, when we set the time delay $\Delta T$ to 4.0ms. It indicates that the sensor measurements $s^1$ and $s^2$ are close to a time-shifted copy of each other. Furthermore, we can also see that the white color regions form a circular strip, meaning that when the motor is placed within this region, the motor magnetic signals collected by the two sensors have a time delay of 4.0ms. The same pattern appear in other figures, where we can see that when we set the time delay to different values, *SER* will have high values within a circular strip region. In other words, by adjusting the time-shifts among different sensors, we can amplify the magnetic signals emitted from certain regions, i.e., the white regions in Figure 7. Note that the signal enhancement pattern is different from the spatial filtering for radio signals, which was highly directional, as reported in [50].

**Delay and Sum Spatial Filtering.** This observation motivates us to leverage the spatial filtering technique, which is a powerful technique for signal enhancement [54, 50]. In this project, we adopt a Delay and Sum Spatial Filtering strategy that is described in Algorithm 1.

---

**Algorithm 1** Delay and Sum Spatial Filtering

---

**Require:** Measurements from $M$ Sensors: $s^i(t)$, $i = 1, ..., M$
**Ensure:** Enhanced Signal $\hat{s}(t)$

1: $s^i(t) \leftarrow IFFT(FFT[s^i(t)]/FFT[n(t)])$
2: **for** $i = 1, i \le M, i + +$ **do**
3:     $\Delta T_i \leftarrow \arg\max_\tau (\sum_{t=0}^{W} s^1(t) * s^i(t + \tau))$
4:     $s^i(t) \leftarrow s^i(t + \Delta T_i)$
5: **end for**
6: $\hat{s}(t) \leftarrow \sum_{i=1}^{M} s^i(t)$

---

The first step is to filter the sensor data and reduce the impact of noise. We have tested multiple filtering algorithms, including notch filtering and bandpass filtering, and found that the adaptive noise cancellation technique [45] is the most effective. In particular, as illustrated in Algorithm 1 line 1, for each input signal $s^i(t)$, we calculate its Fourier transform $FFT[s^i(t)]$. We collect a reference background signal without rotational devices to capture the noise spectrum, denoted by $FFT[n(t)]$, which is correlated solely with environmental interference. To mitigate noise amplification from near-zero values in the noise spectrum estimate, we apply the spectral floor threshold by setting noise frequency bins to 1 when their values are below 1. We then divide the frequency domain signal by the Fourier transform of background noise, and perform an inverse Fourier transform and obtain the filtered signal.

We then estimate the time delays between signals collected at different sensors using the cross correlation-based approach, as described in line 3 of Algorithm 1. In particular, we firstly calculate the cross correlation between two signals, we then select the time delay $\Delta T$ to be the value of $\tau$ that maximizes the signal cross-correlation. Each signal $s^i(t)$ is shifted by its delay $\Delta T_i$ to signal $s^1(t)$ (line 4). Finally, as shown in line 6, we obtain the enhanced signal $\hat{s}(t)$ by summing the time-delayed versions of all the signals.

**Power Spectrum Estimation.** After obtaining the enhanced motor magnetic signal $s(t)$, we will next extract its frequency domain features by computing its power spectrum. Many algorithms are available for estimating the power spectrum, which can be classified into parametric [7, 39] and non-parametric [39] approaches. We select Welch's algorithm, a non-parametric method, due to its simplicity and its ability to control the frequency domain resolution. We use the classical Welch's algorithm described in [6] to extract two power spectrums with two different frequency-domain granularity, to be used in the coarse- and fine-grained rotation speed estimation, as shown in Figure 2. Finally, we apply a logarithmic transformation to the power spectrum and normalize the signal to between the values of 0 and 1.

# 4 PYRAMID POWER SPECTRUM PARSING

Our next task is to detect the presence of motor magnetic signal and identify its frequency from the sensor measurement. The signal detection problem is a well-studied area in communication with decades of research. When the signal quality is good, we can use classical digital signal processing techniques used in the existing magnetic tachometers[27], such as thresholding [20], matched filtering, auto-correlation[30], peak detection [9], and covariance-based algorithms, to detect the motor magnetic signals with high accuracy. However, none of these digital signal processing techniques remain robust to severe signal noises in our data, which occur when we increase the sensing distance, such as the sample signals in Figure 3 (c) and (d), where the motor magnetic signals are overwhelmed by noises.

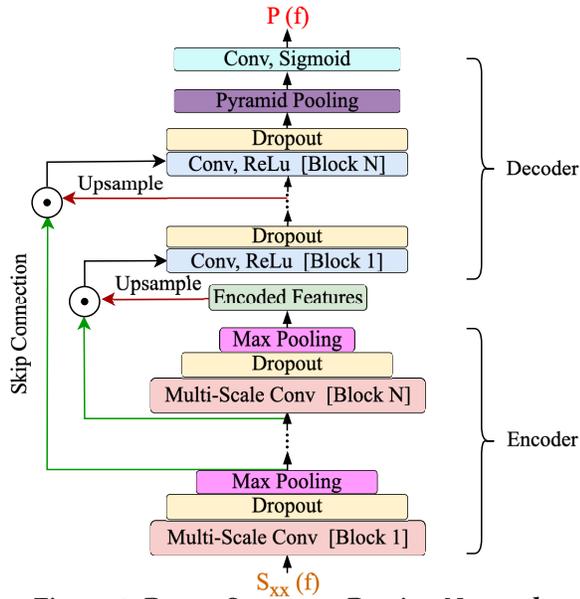

**Figure 8: Power Spectrum Parsing Network**

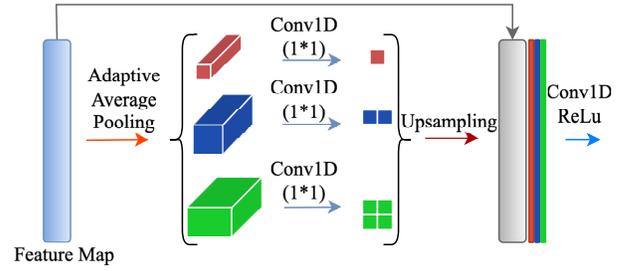

**Figure 9: Pyramid Pooling Module**

In recent years, data-driven approaches have become mainstream techniques for signal detection under noisy conditions [51, 4, 25, 28, 21]. SigDetNet [28], initially developed for detecting wireless communication carrier signals with unknown frequencies, is proposed as a potential algorithm for identifying motor magnetic signals. The SigDetNet is an encoder-decoder network that detects carrier signals based on local and regional contexts using multiple pyramid pooling layers. However, this design is not optimal for capturing the motor magnetic field signal harmonic patterns, which span the global sensor signal spectrum. As a result, in our experiments, SigDetNet produces false detections for noises with high power spectral densities.

To adapt SigDetNet for magnetic signal harmonic detection, we introduce three major improvements: First, to capture the mutual correlations among the power spectrum peaks corresponding to the signal harmonics, we need to significantly increase the reception field of the neural network in order to capture the global correlations. Therefore, we employ a pyramid pooling layer at the top level, which captures the global context information. Second, we use skip connections[55, 48, 46] to enhance feature reusability and improve data training efficiency. Third, we employ multi-scale convolutional layers to capture the diverse harmonic patterns of motor signals. In the following section, we provide a detailed description of the proposed Pyramid Power Spectrum Parsing Network (PPSP Net), depicted in Figure 8.

## 4.1 Architecture for PPSP Net

**Multi-Scale Frequency Feature Extraction.** The PPSP Net comprises downsampling and upsampling modules, where we encode the input features and then decode them to generate the desired output. The input, $S_{xx}(f)$, passes through the multi-scale convolution block [47], followed by the max pooling layer at each level. Compared with the basic convolution blocks used in SigDetNet, the multi-scale convolution layers can detect signal peaks of various sizes and at different resolutions to improve detection accuracy[12]. The convolution layer used in the block consists of 64 filters with a 1*3 kernel size. The max pooling operation with a kernel size of 1*2 is applied to the output of the convolution module before passing on to the next level. There are in total 9 downsampling layers with the same configuration.

In the upsampling module, we aim to reconstruct the dense features learned in the previous phase. At every level of the upsampling phase, we perform three operations: upsampling, concatenation, and convolution. The input from the previous level is upsampled using linear interpolation by a factor of 2 to ensure a matching size for the subsequent concatenation operation. The number of the upsampling section matches the number of the downsampling section to ensure that the output dimensions match those of the input.

**Skip connection.** In our experiments with the SigDetNet, we experienced the gradient degradation problem, meaning that the neural network parameters were not updated properly with the available training data. To address this problem, we introduce the skip connections [55], which involve bypassing intermediate layers by directly connecting the output of one layer to subsequent layers. The skip connections mitigate the gradient degradation issues and improve the reusability of features, allowing the network to leverage fine-grained features learned from previous layers directly.

**Pyramid Pooling.** The Pyramid Pooling layer is used in SigDetNet to fuse the signal features with different reception fields, including local, regional, and global levels [28]. However, unlike SigDetNet that deploys multiple pyramid pooling layers at different levels of the neural network, in PPSP, we strategically deploy the pyramid pooling layer towards the end of the network, as illustrated by the purple block in Figure 8. This way, the the pyramid pooling will fuse the global-level features, the recurring signal harmonics, with the local features, the high amplitudes of the signal peaks in the power spectrum. This design encourages the detection of power spectrum peaks at frequencies that are correlated with each other by integer multiples.

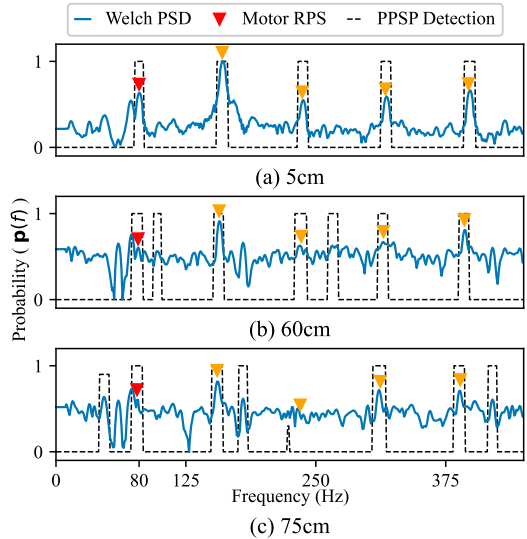

Figure 10: Sample Signal Harmonic Detection Results

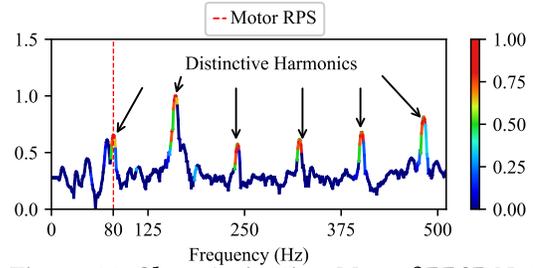

Figure 11: Class Activation Map of PPSP Net

Figure 9 illustrates the internal architecture of a pyramid pooling module. The input to the pyramid pooling layer is the upsampled 1D feature map with 1024 data points. The pyramid pooling module consists of 1) adaptive averaging pooling used to extract features at different scales, local and global, 2) Conv1D layer to reduce the dimensionality and aggregate features, 3) Upsampling operation helps to scale the features to match the size of input feature map for concatenation. We use three average pooling operations on the input feature map with varying kernel sizes of 1*1, 1*2, and 1*4. The output of the pyramid pooling layer is obtained by concatenating the input feature map with the upsampled multi-scale pooled features.

The final convolution module consists of a convolution layer, a Batch Normalization 1D layer and a sigmoid activation function. The output $\mathbf{P}(f)$ will be the probability that a magnetic signal harmonic exists at a frequency $f$.

## 4.2 Model Training

**Data Augmentation.** Next we present our strategies for generalizing the proposed PPSP Net. In practice, it is challenging to collect training data for all possible rotation speeds, because many motors only have discrete rotation speeds, and it is labor-intensive to collect data for all possible speeds. We instead leverage the data augmentation technique to generate training data for different rotation speeds. For each training data sample $s(t)$ with a motor rotation round per second (rps) of $f_0$, we synthesize additional training samples as follows. We resample $s(t)$ by a factor $\alpha$. This way, we obtain a new synthesized signal of $s_\alpha(t)$, with a rotation speed label of $\alpha * f_0$. By adjusting the values of $\alpha$ within the range $[\alpha_{min}, \alpha_{max}]$, we can synthesize a large number of additional training samples.

For MagTach, we collected two sets of data, each from brushed and brushless motors. Each set of real-world data consists of 105 samples, 5 samples collected for every distance varied by multiple of 5 from 5cm to 105cm. There are 420 samples across four sets of real-world data from both types of motors. We then adjust the value of the resampling factor $\alpha$ within the range $[0.1, 2]$ in increments of 0.1.

**Loss Function Selection:** Traditional semantic segmentation or scene-parsing algorithms typically use the cross-entropy functions for loss calculations. However, the cross-entropy function is ineffective in our case because we have significant imbalance in data [29], where the number of positive labels (signal) is much larger than the number of negative labels (noise). Instead, we use the dice loss function instead, which handles imbalances in the label effectively. The dice loss is defined as follows: Dice Loss $= 1 - \frac{2*(\hat{y}*y)}{\hat{y}+y}$. The dice loss' numerator calculates the intersection between the model's prediction and the ground truth labels, while the denominator is just the union of both. A lower value of dice loss indicates more similarity between prediction and ground truth labels.

**Sample Signal Harmonic Detection Results.** We next present sample signal harmonic detection results. We can see that when the signal quality is high, the PPSP Net detects all signal harmonics accurately, as illustrated in Figure 10 (a). Detection errors appear when the signal quality deteriorates. For example, we have false positive detections in Figure 10 (b) at around 100Hz and 260Hz, and in Figure 10 (c) at around 60Hz, 160Hz, and 400Hz. We also have a false negative detection at 240Hz in Figure 10 (c). In Section 5, we will present our strategies to achieve accurate fundamental frequency estimation despite errors in signal harmonic detection.

We use the Class Activation Map (CAM) [33] to visualize and interpret the detection results of PPSP Net. In Figure 11, we plot the contributions of different regions in detecting the first harmonic located at the Motor RPS. We can see that the regions around the other harmonics peak across the entire power spectrum are all making higher contributions, demonstrating PPSP Net's ability to leverage the global context, i.e., the frequencies of other signal harmonics, to assist in predicting the first harmonic.

## 5 ROTATION SPEED ESTIMATION BASED ON BAYESIAN REGRESSION

**Motivations.** When the PPSP network accurately detects signal harmonics, we have multiple options for estimating the rotation speeds, such as the D-frequency extraction [19] and Alias-Elimination [14] algorithms, as reported in the literature. However, unlike active radio systems used in [19] and [14] that produced highly accurate signal harmonic detection, the passive magnetic field used in Magtach has a severe path loss and detection errors begin to emerge at around 50 cm, even after our efforts in signal enhancement and detection. For example, in Figure 10 (b), PPSP produces a false positive detection at 90Hz, and multiple false positive and false negative detections in Figure 10 (c). In this case, deterministic algorithms used in [19] and [14] will produce a wrong rotation speed estimation result. To overcome the uncertainties in signal detection, we propose to leverage the Fuzzy Logic algorithm, which can estimate the motor rotation speeds even when signal harmonic detections are uncertain.

**Fuzzification.** In the fuzzification step, we use $\mathbf{y}(f)$ to denote the likelihood that the magnetic signal's fundamental frequency $f_0$ is close to the frequency $f$, i.e., $|f - f_0| < \Delta f$. $\Delta f$ is a tunable parameter, and we set it to be 1hz in the fuzzification step.

**Fuzzy Fundamental Frequency Estimation Rule.** By definition, the harmonics are integer multiples of the motor signal's fundamental frequency $f_0$. Mathematically, we can describe the correlation using the following equation:

$$\mathbf{y}(f_0) = \beta_{f_0,1} \cdot \mathbf{p}(1) + \beta_{f_0,2} \cdot \mathbf{p}(2)... + \beta_{f_0,M} \cdot \mathbf{p}(M) \quad (4)$$

In this equation, $\mathbf{y}(f_0)$ is positively correlated to $\mathbf{p}(f)$ if $f$ is an integer multiple of $f_0$, and negatively correlated otherwise. We use the Bayesian regression algorithm to obtain the value of $\beta$.

**Coarse-Grained Fundamental Frequency Inference.** A large value of $\mathbf{y}(f)$ is a strong indicator of whether the motor's fundamental frequency $f_0$ is within the range of $[f - \Delta f, f + \Delta f]$. However, there are still random noises that can generate false peaks. To filter the noises out, we will remove a peak $\mathbf{y}(f)$ if $\mathbf{p}(m \cdot f), m = 2, 3, ..., N$ are all zero. A coarse fundamental frequency estimation $f'$ will be selected such that $\mathbf{y}(f')$ is maximized, i.e., $f' = \arg\max_f \mathbf{y}(f)$.

**Fine-Grained Frequency Estimation (Defuzzification).** In the final step, we will obtain a fine-grained estimation of the signal's fundamental frequency. To achieve this goal, we up-sample the sensor measurement $\hat{s}(t)$ by a factor $\gamma$ and recalculate the power spectrum using Welch's method. This way, we improve the frequency domain resolution by a factor of $\gamma$. We then apply a peak finding algorithm on the new power spectrum within the range of $[f' - \Delta f, f' + \Delta f]$ to find the frequency $f_0$ with the largest power spectral density, which will be the final rotation speed estimation. In this project, we have selected $\gamma = 50$, which improves the frequency estimation resolution from $1hz$ ($\Delta f = 1Hz$) to 0.02Hz, i.e., from 60 RPM to 1.2 RPM.

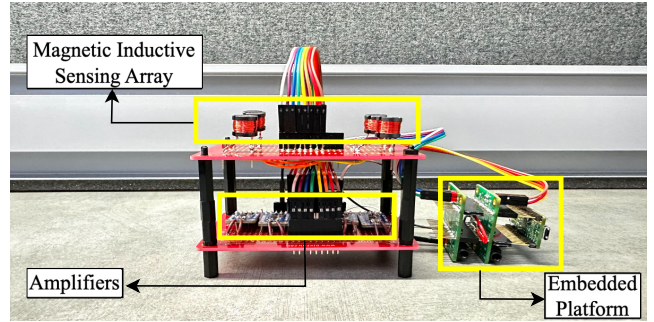

**Figure 12: MagTach System Implementation**

## 6 EVALUATION

### 6.1 Experiment Setup

#### 6.1.1 Prototype System Implementation.
Our implementation of MagTach is illustrated in Figure 12. Four custombuilt inductive sensors are amplified by the MAX9814 chip with a gain of 60dB. The analog signals are digitized using SGTL5000 at 44,100Hz with a 16-bit resolution. The sensor measurement data is stored on the onboard SD card and ready to be transferred to the computer for data processing. The system is driven by a Teensy 4 micro-controller, and is powered by a Lithium battery with a capacity of 2000mAh.

#### 6.1.2 Ground Truth Rotation Speed Acquisition.
To acquire fine-grained ground-truth rotation speeds of a motor, we use the Daedalus CNC Brushless Spindle Motor, a Brushless Direct Current (BLDC) motor, which features embedded speed estimation with a resolution of 0.1 Rotations Per Minute (RPM). At a medium rotation speed of 5000 RPM, this results in a ground-truth speed estimation error of approximately 0.02%. The BLDC motor's rotation speed is controlled by a 0–10,V input DC voltage. To ensure a precise input voltage, we implemented a voltage divider circuit. We also leverage a commercial laser tachometer to measure the rotation speeds of other motors.

#### 6.1.3 Evaluation Metrics.
Similar to prior work on rotation speed estimation systems, we use the root mean absolute error (RMAE) quantify MagTach's estimation accuracy. We use $N$, $\hat{f}$, and $f$ to denote the total number of samples, predicted rotation speed, and rotation speed ground truth, respectively. Then RMAE is calculated as $RMAE = \frac{1}{N} \sum_{i=1}^{N} |\hat{f}_i - f_i| / f_i$.

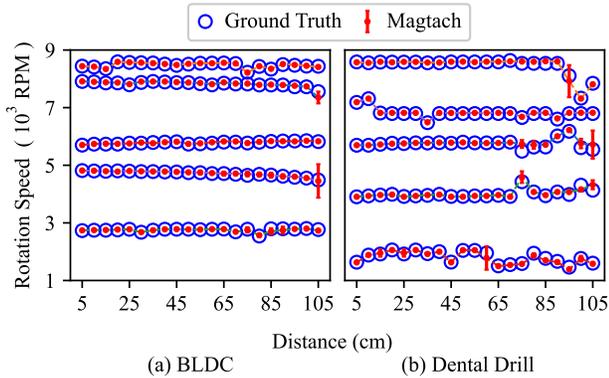

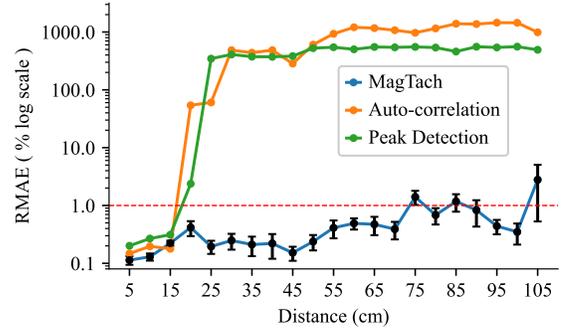

**Figure 13: Sample MagTach Monitoring Results**

*6.1.4 Baseline Algorithms.* 1) The auto-correlation algorithm has been utilized by some magnetic sensing-based rotation speed estimation systems [18, 34]. In essence, this algorithm computes the auto-correlation of the input magnetic inductive sensor measurements and identifies the peak value to determine the period of motor rotation.

2) Peak Detection algorithm [9] identifies the largest peak in the signal's power spectrum, and returns the corresponding frequency as the motor's rotation speed.

*6.1.5 Sample Monitoring Results.* In Figure 13 we present sample rotation speed estimation by varying the distance for the BLDC and Dental drill. We can see that MagTach maintains accurate monitoring accuracy across different RPMs, with a sensing range of up to 105cm.

## 6.2 Micro-Benchmark Experiments

In a series of benchmark experiments, we aim to explore how MagTach performs under different conditions, including different ranges, orientations, and visibility occlusions.

*6.2.1 Rotation Speed Estimation Accuracy vs. Distance.* In this study, we used MagTach to estimate the BLDC motor's rotation speeds at distances from 5cm to 105cm, with a 5cm incremental increment. At each distance, we collected 20 data samples, and each sample was 1 second long. We repeated this experiment for five different motor rotation speeds, ranging between 2000 and 9000 RPM. In total, we collected approximately 2100 samples, which was 0.5 hours of data.

We provide an overview of the results for BLDC in Figure 13, where each blue circle represents a data sample's rotation speeds (Y-axis) and distances (X-axis). Overall, we can see that MagTach estimates the speeds most of the time accurately. We can also see small speed estimation errors when the distances are above 75cm, and bigger errors at 105cm.

**Figure 14: RMAE vs. Distance**

In Figure 14, we present the aggregated results to illustrate MagTach's performance over distance. The RMAE of MagTach is 0.21%, 0.34% and 0.54% when the distance is below 50cm, 70cm, and 100cm, respectively. When the distance is 75cm and 100cm, the rotation speed estimation error grows, but the RMAE remains around or below 1.1% while the mean RMAE below 100cm was 0.54%. At a distance of 105cm, the rotation speed estimation error is 2.3%. Magtach outperforms the laser tachometers, which achieve an error rate of around 0.5% with a range of up to 70cm [56]. However, a laser tachometer requires attaching reflective tags to the rotating components, and the monitoring accuracy degrades when either the hand-held laser gun or the target device is shaken, which is a common situation. In comparison, MagTach does not require careful alignments between the sensor and the targets, and small shaking has minimal impacts on the monitoring accuracy.

We also tested the performance of baseline signal period finding algorithms, the Auto-Correlation based, and the Power Spectrum Peak Finding approach, as described in Section 6.1.4. As illustrated in Figure 14, Auto-correlation exhibits errors as low as 0.3% up to 15 cm but increases rapidly between 20 cm and 105 cm, respectively. Peak detection follows a similar pattern, with errors ranging from 0.3% to 0.4% up to 15 cm and then rapidly increasing afterward. Between 20 cm and 40 cm, these algorithms tend to estimate based on the second or third harmonics rather than the fundamental frequency, which causes 100% or 200% of errors. Beyond 45 cm, the error trend stabilizes as they begin predicting high-frequency noise due to signal distortions. The baseline algorithms' low monitoring accuracy at distances above 20cm is due to the rapidly diminishing strength of the motor magnetic field.

*6.2.2 Rotation Speed Estimation Accuracy vs. Orientations.* In this experiment, we rotated the BLDC's cardinal orientation by 60-degree increments and collected 20 samples at each orientation. We present our results in Figure 15.

In this figure, we can see that the rotation estimation RMAE have low values between 0.08% and 0.18% when the motor's orientation is between $-120°$ and $120°$. The RMAE increases to 0.45% when facing the sensors from the backside, which is 180°. This experiments shows that MagTach remains accurate even when there is no accurate alignments between the sensor and the rotating device.

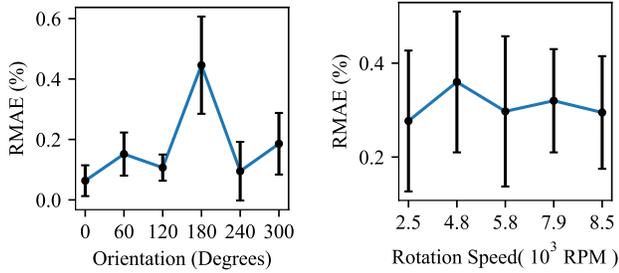

**Figure 15: RMAE vs. Orientation**

**Figure 16: RMAE vs. Rotation Speed**

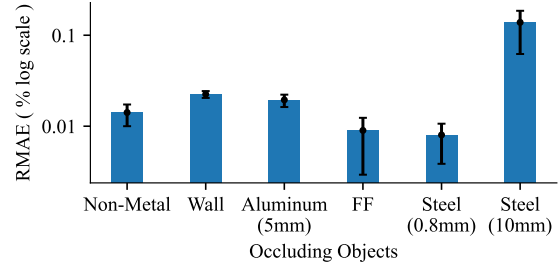

**Figure 17: RMAE vs. Occluding Objects**

*6.2.3 Rotation Speed Estimation Accuracy vs. Target Speeds.* In Figure 16, we present the MagTach's speed estimation accuracy when the target device has rotating speeds ranging from 2700 to 8500 RPM. The data samples are collected when the distances between the sensor and the motor are between 5 and 75 cm. We can see that MagTach maintains a high speed estimation accuracy across a wide range of speeds. In particular, the errors are below 0.4% over the RPM ranges, with the lowest of 0.28% at 2700 RPM and the highest at 0.38% for 4800 RPM. The estimation variation remains consistent between ± 0.1% to 0.2% with the highest variation of ± 0.19% for 5800 RPM and the lowest being ± 0.1% for 7900 RPM.

*6.2.4 RMAE vs. Occluding Objects.* In Figure 17, we compare the RMAE when there are different occluding objects. In this experiment, we aim to study MagTach's ability to estimate rotation speeds despite occlusions. We have tested multiple types of obstacles, including 1) non-metal obstacles such as plastic, cardboard boxes, and glass; 2) a wooden wall with a thickness of around 15cm; 3) a piece of $30cm \times 30cm$ Faraday Fabric (FF) that wraps the motor [1]; 4) an Aluminum baking sheet with a thickness of 5mm; 5) a stainless steel (type 304 stainless steel) sheet with a thickness of 0.8mm; and 6) a stainless steel pot with a thickness of 10mm that encloses the motor. After occluding the motor with each material, we placed the sensor at a distance of 25cm and collected 20 samples.

In Figure 17, we can see that when the device is enclosed by non-metal objects, including plastic, cardboxes, and glass, the rotation speed estimation error is at a low level of 0.2%.

---

[1]Faraday fabric is designed to block radio frequency signals. The fabric we use consists of three layers: polyester fiber, copper mesh, and nickel.

The 15cm-thick wall raises the RMAE to 0.5%. We further discover that thin metals, which can block out high frequency radio, doesn't affect MagTach. In particular, MagTach acheives RMAE of 0.4%, 0.1%, and 0.09% when the occluding objects are 5mm-thick hard anodized aluminum, Faraday fabric, and 0.8mm-thick stainless steel sheets, respectively. When we enclose the motor with a 10mm-thick pot, the RMAE increases significantly to 18%.

MagTach's high monitoring accuracy despite obstacles is due to the unique properties of the motor magnetic signals. The motor's magnetic signals have high skin depth due to their low frequencies, allowing them to penetrate thin metal objects and many non-conductive materials.

## 6.3 Generalizability of MagTach

*6.3.1 RMAE vs. Unseen Devices.* As illustrated in Figure 1, we used MagTach to monitor the rotation speeds of multiple different devices. In this subsection we will evaluate MagTach's ability to generalize to unseen devices. In this experiment, we collected 20 data samples from each of the following devices.

1) **Dental Drill:** It is usually referred to as Dental Micromotor [32]. This tool is used without a rotating blade but just with the rotating spindle attachment, which has a diameter of around 6mm and a length of 15mm. The rotation speed can be controlled and ranges from 0 to 9,000 RPM.

2) **Hand Drill:** This Black-Decker drill[5] is driven by a brushed Direct Current (DC) motor. The chuck size is $3/8^{th}$ of an inch and is rated maximum at 650 RPM.

3) **Roller Pump:** This is also called a peristaltic pump, commonly used to transfer various fluids in medical applications. This device doesn't have any exposed rotating blades and operates at 5000 RPM when powered by a DC 12V supply.

4) **USB Mini Fans:** We tested three USB fans that use brushed DC motors and operate on 5V DC input. One of the fans has 3-speed settings of 1000, 2100, and 4500 RPM, while the other two fans operate at 5800 RPM and 7100 RPM, respectively.

5) **Electric Toothbrush (ET):** We tested on Oral-B electric toothbrush. It does not have external rotating blades, so we rely on the manufacturer's specification of 7500 RPM as the ground-truth rotation speed.

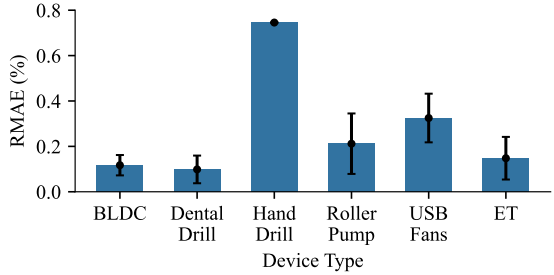

**Figure 18: RMAE vs. Device Type**

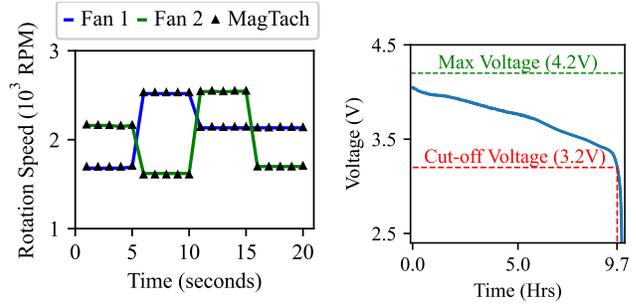

**Figure 19: Concurrent De-** **Figure 20: Battery Voltage**
**vices Monitoring** **vs. Time**

We used one set of data from both the BLDC motor and a USB fan to train the PPSP Net algorithm and then applied the trained model to monitor the rotation speeds of all the devices. Figure 18 presents the RMAE results for MagTach. The results show that MagTach maintains high monitoring accuracy for unseen motors. For example, the monitoring accuracies for the Dental Drill, Roller Pump, and Electric Toothbrush (ET) are 0.1%, 0.22%, and 0.16%, respectively, comparable to those of the BLDC motor and USB fan. The hand drill's maximum RPM estimation error is slightly higher at 0.7%, primarily due to its unstable rotation speed over time. The average estimation error for unseen devices is 0.29%, demonstrating MagTach's ability to generalize its monitoring model to different motors without prior training data.

*6.3.2 RMAE vs. Motor Design.* Next, we evaluate MagTach's generalizability to different motor designs, specifically brushed and brushless DC motors. In this evaluation, we train MagTach using data from either the brushed motor (BR) only or the brushless DC (BLDC) motor only. We then assess the monitoring accuracy of these two models on both types of motor data, with the results presented in Table 1.

| Test<br>Train | BR | BLDC |
|---|---|---|
| BR | 0.32% | 0.68% |
| BLDC | 0.52% | 0.11% |

**Table 1: Mean RMAE for Generalization Validation**

We observe that while MagTach maintains relatively high accuracy under all four scenarios ($RMAE \leq 0.68\%$), it performs slightly worse compared to laser tachometers ($RMAE \approx 0.4\%$). Notably, there are observable degradations in monitoring accuracy, around 0.4%, when generalizing MagTach to different motor designs (increasing from 0.32% to 0.68% in the first row and from 0.11% to 0.52% in the second row of Table 1). Nonetheless, our experiments show that MagTach can generalize well to four unseen motors using data from just two motors: one BLDC motor and one brushed USB fan.

### 6.4 Concurrent Devices Monitoring

In this study, we evaluated MagTach's ability to monitor multiple devices concurrently. For this experiment, we placed two fans at an equal distance of 30,cm from MagTach. As shown in Figure 19, the green and blue lines represent the actual rotation speeds of the two fans—1600, 2100, and 2600 RPM—over time. MagTach's rotation speed estimation results are indicated by triangle markers in Figure 19. The results show that MagTach closely tracked the speed changes of both fans. Notably, the RMAE (Relative Mean Absolute Error) of the rotation speed estimation for the two fans was just 0.2%.

### 6.5 Energy Consumption of MagTach

In this experiment, we aim to evaluate MagTach's energy consumption rate. As discussed in Section 6.1.1, our current prototype implementation includes four inductive sensing channels. Each channel continuously records the magnetic signal at a 44.1 kHz sampling rate, and the four-channel data is recorded on a local SD card. The entire system is powered by a 2000mAh battery. To assess the system's energy consumption rate when recording data continuously, we measure the battery's voltage every second, and the results are shown in Figure 20.

We can see that when fully charged, the battery has a voltage of 4.1V. Due to MagTach's operation, the battery's voltage drops gradually. It takes 9.7 hours for the battery to reach its cut-off voltage of 3.2V, and the system cease working. The total energy of the battery that was consumed in this process can be expressed as $2Ah * 3.7V = 7.4Wh$. Dividing the total duration of 9.7 hours, we estimate that the energy consumption rate of MagTach is 760 mili-Watt.

### 6.6 Discussion

**Mechanical Rotating Devices.** In this project, we have focused exclusively on the rotation speed estimation of electric machines, which use electromagnetic force to generate rotation torque. Nevertheless, there are many mechanical rotating machines that do not use electromagnetic force, such as Internal Combustion Engines (ICE) and wind mills. To extend MagTach to these types of devices without any ferrite components, we can potentially attach passive magnetic tags on the rotating components, and use the magnetic fields to monitor the rotation speeds.

**Interference from Nearby Devices.** In Section 6.4, we demonstrated that MagTach can monitor two rotating devices concurrently. One potential challenge is whether MagTach can still accurately estimate the speed of a specific device in a crowded environment with many rotating devices and electronic components. Through multiple experiments, we observed that while MagTach's sensing is not particularly directional, it is highly sensitive to distance. This is because the strength of the low-frequency magnetic field is inversely proportional to the fourth power of distance. As a result, we can easily monitor the device of interest by holding MagTach at a closer distance—preferably within 30cm—where the target device's signal will dominate due to its relative strength. In our tests, we found it straightforward to select and monitor the desired device.

# 7 RELATED WORK

**Optical sensing systems**, such as those using lasers and cameras[16, 15, 35, 56], have gained popularity due to their convenient form factor and accuracy. Laser-based systems[53] are widely available on the market. Yet, they require attaching reflective tags on the rotating objects, which are hard to achieve in many applications[56]. Furthermore, the laser tachometers require precise alignment between the sensor and the small reflective tags, making handheld usage challenging. Camera-based systems, as noted in [16, 15, 56], eliminate the need for reflective tags using video processing algorithms. These systems also provide enhanced robustness o the users' hand shakiness, allowing for handheld rotation speed monitoring.

However, all optical sensing-based systems require line-of-sight (LoS) visibility, making them ineffective in challenging environments with significant obstructions, such as warehouses and factories, in low-light conditions, or when the target is fully covered, as in the case of a liquid pump. In comparison, MagTach estimates rotation speed through various obstacles, including walls and even thin metals. Furthermore, MagTach is effective in monitoring rotation speeds exceeding 9,000 RPM, significantly higher than the 6,000 RPM achieved by EVTach. These new features are achieved while maintaining a similarly convenient form factor (handheld) and sensing range ($\approx 1m$) as optical sensing based systems.

**RF-Based Solutions.** Active RF sensing technology emits radio signals and measures those reflected by rotating blades, enabling impressive sensing ranges for multiple systems [13, 19, 26, 14]. For example, RFTacho, mRotate, and WiRotate have sensing ranges of 0.6, 2.5, and 3m, respectively. However, these systems have different abilities for through-obstacle sensing, which depends on the signal characteristics. In particular, the mmWave-based [26, 14, 44] and WiFi-based

| Method | Non LoS | Handheld Sensing | Sensing Range | Max. RPM |
|---|---|---|---|---|
| Laser[53] | × | ✓ | $0.2m$ | 100k |
| Magnetic[3] | ✓ | × | $0.025m$ | 2.4k |
| EVTach[56] | × | ✓ | $0.7m$ | 6k |
| RFTacho[19] | ✓ | × | $0.6m$ | 6.5k |
| mRotate[14] | × | ✓ | $2m$ | 11k |
| **MagTach** | ✓ | ✓ | **$1m$** | **9k** |

**Table 2: SotA Rotation Speed Estimation Methods**

[13] systems are unable to sense through obstacles accurately, while the OAM Wave-based system can sense through obstacles including plastics and glass [19], no tests were conducted about the heavier obstacles like woods or meal. Furthermore, all three RF-based systems require fixed setups, limiting their handheld usage. In contrast, MagTach, with a sensing range of up to 1 meter, employs a purely passive system design, eliminating the need for active signal emission. This passive approach provides significant advantages in terms of system form factor, energy efficiency, and cost, enabling the development of a mobile solution comparable to mass-market laser tachometers. Besides, MagTach offers a superior ability for penetrating obstacles, which enables new application scenarios, including through wall monitoring and monitoring for embedded components, such as medical pumps.

**Mechanical and Magnetic Solutions.** Vibration-based rotation speed estimation techniques are widely used in industrial applications [17, 38, 24, 37]. These operate in proximity and are not robust to ambient vibrational noises. Existing magnetic tachometers [42, 34, 1, 49, 2, 43, 31] are a noninvasive solution, where they produce signals with frequencies proportional to the rotation speed. These methods require prior knowledge about the rotating machines, such as the number of poles inside the motor, take significant efforts to deploy, and have short sensing range. In comparison, MagTach achieves accurate rotation speed estimation at a distance of 100cm in a convenient handheld form-factor.

**Magnetic Inductive Sensing.** In recent years, the magnetic inductive sensing technology has attracted significant research interest [11, 22]. In [11], the inductive sensor has been used to sniff the electromagnetic emission from visible light communication through the wall at a distance of 6 meters. In comparison, the signals MagTach focuses have about a thousand times lower frequencies, which result in weaker induced signals. To mitigate the challenge, MagTach uses ferrite cores instead of the the air core and the spatial filtering technique to enhance the signal quality.

# 8 CONCLUSION AND FUTURE WORK

In this work, we propose MagTach, the first electromagnetic-based sensing system for measuring the rotation speed of a device from a distance of up to 1 meter. Multiple modules, including the spatial-filtering-based signal enhancement algorithm, the Pyramid Power Spectrum Parsing Network, and the fuzzy-logic-based signal fundamental frequency estimation, were developed. Through extensive evaluations, MagTach achieved state-of-the-art performance with mean estimation errors of 0.21%, 0.34%, and 0.54% at distances of 50cm, 70cm, and 100cm, respectively. MagTach remained robust when devices changed orientation, operated at speeds between 650 to 9,000 RPM, and were tested across six types of devices, including those without external rotating components. MagTach demonstrates strong capabilities for sensing through occluding obstacles, including a 15cm-thick wall and 0.8mm-thick stainless steel sheets, all in a compact handheld form-factor with a low energy consumption.